\documentclass[usenatbib, aps, pra, amsmath, amssymb, 11pt, final, tightenlines, twoside, twocolumn,nofloats, nofootinbib, superscriptaddress, showkeys, showkeywords]
{revtex4}

\usepackage[T2A]{fontenc}
\usepackage[utf8x]{inputenc}
\usepackage[english]{babel}
\usepackage{graphicx}
\usepackage{dcolumn}
\usepackage{bm}
\usepackage{longtable}
\usepackage[usenames]{color}

\def\squareforqed{\hbox{\rlap{$\sqcap$}$\sqcup$}}

\def\sq{\ifmmode\squareforqed\else{\unskip\nobreak\hfil
\penalty50\hskip1em\null\nobreak\hfil\squareforqed
\parfillskip=0pt\finalhyphendemerits=0\endgraf}\fi}

\def\utw{\smash{\rlap{\lower5pt\hbox{$\sim$}}}}

\def\udtw{\smash{\rlap{\lower6pt\hbox{$\approx$}}}}

\def\diameter{{\ifmmode\mathchoice
{\ooalign{\hfil\hbox{$\displaystyle/$}\hfil\crcr
{\hbox{$\displaystyle\mathchar"20D$}}}}
{\ooalign{\hfil\hbox{$\textstyle/$}\hfil\crcr
{\hbox{$\textstyle\mathchar"20D$}}}}
{\ooalign{\hfil\hbox{$\scriptstyle/$}\hfil\crcr
{\hbox{$\scriptstyle\mathchar"20D$}}}}
{\ooalign{\hfil\hbox{$\scriptscriptstyle/$}\hfil\crcr
{\hbox{$\scriptscriptstyle\mathchar"20D$}}}}
\else{\ooalign{\hfil/\hfil\crcr\mathhexbox20D}}%
\fi}}

\setcitestyle{authoryear, round, aysep={,}, citesep={;}}
\begin{document}


\title{RECURRENT SYMBIOTIC NOVA T CORONAE BOREALIS BEFORE OUTBURST}

\author{\firstname{N.~A.}~\surname{Maslennikova}}
\email{maslennikova.na16@physics.msu.ru} \affiliation{Sternberg Astronomical
Institute, Lomonosov Moscow State University, Moscow, 119234
Russia} \affiliation{Faculty of
Physics, Lomonosov Moscow State University, Moscow, 119191 Russia}

\author{\firstname{A.~M.}~\surname{Tatarnikov}}
\affiliation{Sternberg Astronomical Institute, Lomonosov Moscow
State University, Moscow, 119234 Russia} \affiliation{Faculty of
Physics, Lomonosov Moscow State University, Moscow, 119191 Russia}

\author{\firstname{A.~A.}~\surname{Tatarnikova}}
\affiliation{Sternberg Astronomical Institute, Lomonosov Moscow
State University, Moscow, 119234 Russia}

\author{\firstname{A.~V.}~\surname{Dodin}}
\affiliation{Sternberg Astronomical Institute, Lomonosov Moscow
State University, Moscow, 119234 Russia}

\author{\firstname{V.~I.}~\surname{Shenavrin}}
\affiliation{Sternberg Astronomical Institute, Lomonosov Moscow
State University, Moscow, 119234 Russia}

\author{\firstname{M.~A.}~\surname{Burlak}}
\affiliation{Sternberg Astronomical Institute, Lomonosov Moscow
State University, Moscow, 119234 Russia}

\author{\firstname{S.~G.}~\surname{Zheltoukhov}}
\affiliation{Sternberg Astronomical Institute, Lomonosov Moscow
State University, Moscow, 119234 Russia} \affiliation{Faculty of
Physics, Lomonosov Moscow State University, Moscow, 119191 Russia}

\author{\firstname{I.~A.}~\surname{Strakhov}}
\affiliation{Sternberg Astronomical Institute, Lomonosov Moscow
State University, Moscow, 119234 Russia}

\begin{abstract}

The results of photometric and spectral observations of T~CrB obtained in a wide range of wavelengths in 2011--2023 are presented. We use the near-IR light curves to determine a new ephemeris $JD_{min} = 2455828.9 + 227.55 \times E$ for the times of light minima when the red giant is located between the observer and the hot component. The flux ratio H$\alpha$/H$\beta$ varied from $\sim 3$ to $\sim 8$ in 2020--2023, which may be due to a change in the flux ratio between the X-ray and optical ranges. It is shown that the value of H$\alpha$/H$\beta$ anticorrelates with the rate of accretion onto the hot component of the system. Based on high-speed follow-up observations obtained on June 8, 2023, we detected a variability of the He\,II $\lambda 4686$ line with a characteristic time-scale of $\sim 25$~min, the amplitude of variability in the $B$-band was $\sim 0\,.\!\!^{\rm m}07$. Simulations of the near-IR light curves accounting for the ellipsoidal effect allowed us to obtain the parameters of the binary system: the Roche lobe filling factor of the cool component $\mu=1.0$, the mass ratio $q=M_{cool}/M_{hot} \in [0.5, 0.77]$, the orbital inclination $i \in [55^\circ, 63^\circ]$. A comparison of the light curve obtained in 2005-2023 with the 1946 outburst template made it possible to predict the date of the upcoming outburst - January 2024.

\end{abstract}

\maketitle

\section{INTRODUCTION}

T~CrB is a famous symbiotic recurrent nova. Throughout the history of observations, T~CrB erupted as a nova twice, in 1866 and in 1946, at maximum light becoming brighter than $2^{\rm m}$. It is expected to flare up again in the near future (Schaefer, 2023).

T CrB is a symbiotic binary~-- a system consisting of a red giant and a white dwarf. Due to the presence of a high-luminosity star in the system (the cool component is a M4\,III star), the amplitude of outburst is small compared to that of classical novae which is about $8^{\rm m}$. Nevertheless, the outburst itself is completely similar to the outburst of a classical nova as it develops on the surface of a white dwarf.

According to Fekel et al. (2000), the orbital period of T~CrB is approximately equal to $P_{orb}=227.6$\,d. At the same time, no eclipses are observed in the system (Selvelli et al., 1992). The light curves of T~CrB in the optical and near-IR ranges demonstrate the presence of periodic variations that occur with a period of $0.5P_{orb}$ and are related to the ellipsoidal shape of the cool component. The amplitude of this effect indicates that the cool component completely fills its Roche lobe (Shahbaz et al., 1997).

In addition to regular long-term variability, T~CrB exhibits irregular brightness variations in the short-wavelength range (Zamanov and Bruch (1998), Zamanov et al. (2004), Zamanov et al. (2005), Minev et al. (2023), etc.) with a characteristic time-scale of tens of minutes and an amplitude up to $\sim 0\,.\!\!^{\rm m}5$ in the $U$ band (by an analogy with cataclysmic variables, this type of variability of symbiotic stars is called flickering). Such variability was also registered by Maslennikova et al. (2023) during spectral observations both for the continuum and emission line fluxes. It is associated with the presence of an accretion disk around the hot component of the system. The accretion rate estimates made by different authors indicate that the accretion rate can significantly change with time. A change in the accretion rate affects the luminosity of the accretion disk, causing changes in the overall spectral energy distribution (SED) of the system which are noticeable at wavelengths shorter than 4500\,\AA.

In 2015 T~CrB entered a so-called super-active phase (Ilkiewicz et al. (2016), Munari (2023)). By 2016 the Balmer lines fluxes had increased by more than an order of magnitude, strong lines of He\,I and He\,II had appeared, the mean $B$ brightness had increased by $1^{\rm m}$, the amplitude of flickering had declined. According to Munari (2023), this phase lasted until about the middle of 2023 with a maximum in April 2016.

The analysis of the complete set of photometric data obtained for T~CrB from 1855 till 2023 showed that the light curves before, during and after the 1866 and 1946 outbursts were very similar. This fact allowed Schaefer (2023) to create a template of light change and to predict a date of $2025.5\pm1.3$ for the next outburst of T~CrB.

We aim to investigate the photometric and spectral variability of T~CrB at the pre-outburst phase and to revise the date of the upcoming eruption.

\section{OBSERVATIONS}

Our spectroscopic observations of T~CrB were carried out at the 2.5-m telescope of the Caucasian mountain observatory of the Sternberg astronomical institute of the Moscow State University (CMO SAI MSU) with the Transient Double-beam Spectrograph (TDS Potanin et al., 2020). During the observations, a 1$''$ slit was used, which provides the spectral resolution $R=1300$ in the short-wave channel (the so-called $B$-channel, wavelength range 3600--5770\,\AA), and $R=2500$ in the long-wave channel ($R$-channel, wavelength range 5700--7400\,\AA). The slit in the TDS was oriented in the direction of the zenith angular distance to minimize atmospheric dispersion. Standard A0V stars were used to correct for tellurics. The observation log is shown in Table~\ref{tab:observation}. In addition to obtaining individual spectra, a 2-hour set of high-speed follow-up observations was carried out on June 8, 2023, when the spectra were recorded continuously (with breaks of 20~s for reading) with exposures of 100\,s in the $B$-channel and 30\,s in the $R$-channel. Standard stars were selected at airmasses close to that of the object, and were observed before the start and immediately after the end of follow-up observations. When calibrating the spectra, the airmass of the standard star was reduced to the airmass of the currently processed T~CrB spectrum.

The obtained spectra were processed according to the algorithm described in Potanin et al. (2020). The spectra were wavelength-calibrated using the emission spectrum of a gas-discharge Ne-Kr-Pb hollow cathode lamp (HCL), corrections for vignetting and non-uniform slit illumination were calculated using a continuum lamp. To check the quality of the conversion of the observed fluxes to the absolute ones, we used the closest in time $V$-band photometric observation from the AAVSO database (Kloppenborg, 2023). All the spectra were reduced to the Solar system barycenter and corrected for interstellar extinction with $E(B-V)=0.15$ (Selvelli et al., 1992). The spectra were processed using self-developed \textsc{python} scripts.

\begin{table*}[t]
\vspace{6mm}
\centering
\caption{Log of the TDS observations for T~CrB}\label{tab:observation} 
\vspace{5mm}\begin{tabular}{l|c|c|c|c}

\hline
Date & Exposure in $B$-channel & Exposure in $R$-channel & FWHM & phase \\
\hline
2020-03-22 & 3 $\times$ 100~s & 4 $\times$ 30~s & 1.5$''$ & 0.63\\
2020-08-25 & 48 $\times$ 150~s & 142 $\times$ 30~s & 1.8$''$ & 0.32\\
2022-01-05 & 8 $\times$ 100~s & 8 $\times$ 30~s & 1.5$''$& 0.50\\
2022-12-09 & 6 $\times$ 100~s & 7 $\times$ 30~s & 0.9$''$& 0.99\\
2023-01-24 & 4 $\times$ 100~s & 6 $\times$ 30~s & 0.9$''$& 0.19\\
2023-01-27 & 2 $\times$ 100~s & 2 $\times$ 30~s & 1.7$''$& 0.21\\
2023-02-26 & 3 $\times$ 100~s & 4 $\times$ 30~s & 1.1$''$& 0.34\\
2023-03-09 & 2 $\times$ 100~s & 4 $\times$ 30~s & 1.2$''$& 0.39\\
2023-04-01 & 3 $\times$ 100~s & 5 $\times$ 30~s & 1.0$''$& 0.49\\
2023-04-25 & 2 $\times$ 100~s & 3 $\times$ 30~s & 1.4$''$& 0.59\\
2023-05-04 & 2 $\times$ 100~s & 3 $\times$ 30~s & 1.6$''$& 0.63\\
2023-05-24 & 4 $\times$ 100~s & 4 $\times$ 30~s & 1.4$''$& 0.72\\
2023-05-26 & 2 $\times$ 100~s & 3 $\times$ 30~s & 1.5$''$& 0.73\\
2023-05-27 & 2 $\times$ 100~s & 3 $\times$ 30~s & 1.1$''$& 0.75\\
2023-05-30 & 2 $\times$ 100~s & 3 $\times$ 30~s & 1.5$''$& 0.75\\
2023-06-08 & 67 $\times$ 100~s & 153 $\times$ 30~s & 1.0$''$& 0.79\\
2023-06-21 & 2 $\times$ 100~s & 3 $\times$ 30~s & 1.0$''$& 0.84\\
2023-06-24 & 2 $\times$ 100~s & 3 $\times$ 30~s & 0.9$''$& 0.86\\
2023-06-26 & 3 $\times$ 100~s & 6 $\times$ 30~s & 1.2$''$& 0.86\\
2023-07-04 & 2 $\times$ 100~s & 3 $\times$ 30~s & 1.0$''$& 0.90\\
2023-07-14 & 2 $\times$ 100~s & 4 $\times$ 30~s & 1.5$''$& 0.94\\
\hline
\end{tabular}\\
\end{table*}

The $B$-band photometric follow-up observations of T~CrB were performed on June 8, 2023 with the RC-600 telescope of the CMO SAI MSU (Berdnikov et al., 2020). The campaign lasted for nearly 2 hours, a total of 348 images with an exposure time of 15~s were obtained. After the initial standard data reduction procedures performed (bias and dark-subtraction and flat-fielding), a differential aperture photometry was applied using the \textsc{MaxIm DL} software package. The comparison stars were chosen among the field stars of comparable brightness.

The infrared (IR) observations were carried out on the 1.25-m telescope of the Crimean astronomical station (CAS) of SAI MSU in 2011--2023 with the one-channel InSb-photometer (Shenavrin et al., 2011), which $JHKLM$ photometric system is close to that of Johnson (1965). The star BS\,5947 ($J=2\,.\!\!^{\rm m}09$, $H=1\,.\!\!^{\rm m}60$, $K=1\,.\!\!^{\rm m}30$, $L=1\,.\!\!^{\rm m}12$, $M=1\,.\!\!^{\rm m}35$) was used as a comparison star. We present the $JK$ photometry in Table~\ref{tab:IR_observation} and the $JHKLM$ photometry in Table~\ref{tab:JHKLM_observations}. The brightness uncertainties are $0\,.\!\!^{\rm m}02$ for $JHKL$ and $0\,.\!\!^{\rm m}05$ for $M$.

\begin{table*}[t]
\centering
\caption { $JK$ photometry for T~CrB}\label{tab:IR_observation} 
\begin{tabular}{c|c|c|c|c|c|c|c|c|c|c|c}
\hline
JD& $J$,  & $K$, &JD & $J$, & $K$, &JD & $J$, & $K$,& JD& $J$,  & $K$,\\
(-2400000) & mag & mag&(-2400000) & mag & mag&(-2400000) & mag & mag&(-2400000) & mag & mag\\
\hline
55760.3&5.88&4.70&56848.3&6.04&4.84&57976.3&6.00&4.81&58963.4&5.83&4.68\\ 
55782.3&5.88&4.72&56871.3&5.98&4.80&58213.5&6.00&4.82&58980.4&5.89&4.72\\ 
55783.3&5.90&4.72&56880.3&5.92&4.76&58231.5&5.98&4.79&59020.4&6.01&4.81\\ 
55783.3&5.90&4.72&56886.2&5.88&4.74&58252.4&5.88&4.72&59036.4&5.91&4.74\\ 
55794.2&5.93&4.74&56886.2&5.91&4.74&58303.4&5.91&4.72&59067.3&5.83&4.67\\ 
56005.6&5.87&4.72&56916.2&5.86&4.70&58334.3&5.96&4.76&59073.3&5.82&4.68\\ 
56033.5&5.98&4.76&56941.2&5.94&4.76&58348.3&5.94&4.74&59288.6&5.83&4.66\\ 
56058.4&6.01&4.81&57103.6&5.93&4.76&58519.7&5.87&4.71&59306.5&5.83&4.67\\ 
56080.4&5.96&4.77&57124.5&5.84&4.68&58534.6&5.93&4.76&59326.4&5.91&4.75\\ 
56089.4&5.88&4.70&57227.3&5.89&4.70&58546.6&5.94&4.78&59376.4&5.93&4.77\\ 
56110.3&5.88&4.68&57261.3&5.84&4.68&58556.6&5.94&4.77&59412.3&5.82&4.69\\ 
56147.3&5.98&4.77&57268.3&5.88&4.72&58561.6&5.92&4.76&59451.3&5.94&4.76\\ 
56409.5&6.00&4.80&57485.5&5.78&4.62&58598.5&5.86&4.70&59652.6&5.86&4.70\\ 
56431.4&5.91&4.73&57525.5&6.00&4.80&58617.4&5.81&4.67&59697.5&5.94&4.77\\ 
56438.4&5.88&4.71&57540.4&5.94&4.75&58627.4&5.84&4.71&59712.4&5.92&4.75\\ 
56442.4&5.86&4.71&57557.3&5.87&4.72&58630.4&5.85&4.69&59741.4&5.86&4.69\\ 
56467.4&5.92&4.72&57566.4&5.88&4.68&58632.4&5.84&4.70&59768.3&5.83&4.68\\ 
56471.4&5.90&4.72&57583.4&5.84&4.65&58636.4&5.90&4.70&59776.4&5.87&4.73\\ 
56485.3&5.94&4.75&57591.3&5.82&4.66&58653.3&6.00&4.79&59986.6&5.81&4.67\\ 
56492.4&5.94&4.75&57620.3&5.92&4.75&58686.3&5.96&4.76&60028.5&5.98&4.81\\ 
56498.3&6.00&4.78&57644.2&5.95&4.75&58706.3&5.86&4.71&60049.5&5.99&4.82\\ 
56517.3&6.03&4.80&57801.6&5.84&4.68&58721.2&5.83&4.70&60068.4&5.90&4.73\\ 
56517.3&6.01&4.82&57812.6&5.82&4.65&58735.2&5.83&4.68&&&\\ 
56517.3&--&4.81&57834.5&5.84&4.68&58875.6&5.91&4.76&&&\\ 
56694.7&5.88&4.72&57846.5&5.90&4.71&58889.6&5.98&4.81&&&\\ 
56739.6&6.00&4.82&57878.4&5.92&4.73&58909.5&5.95&4.78&&&\\ 
56769.5&5.91&4.75&57908.4&5.87&4.69&58921.6&5.92&4.75&&&\\ 
56824.4&5.92&4.76&57971.3&5.92&4.73&58949.5&5.82&4.67&&&\\ 

\hline
\end{tabular}
\end{table*}

\begin{table*}[t]
\caption {$JHKLM$ photometry for T~CrB} \label{tab:JHKLM_observations}
\centering
\begin{tabular}{c|c|c|c|c|c}
\hline
JD (-2400000) & $J$, mag & $H$, mag & $K$, mag & $L$, mag & $M$, mag \\
\hline
55760.3 & 5.88 & 5.03 & 4.70 & 4.36 & -- \\
56089.4 & 5.88 & 5.04 & 4.70 & -- & -- \\
57976.3 & 6.00 & 5.15 & 4.81 & 4.44 & -- \\
58303.4 & 5.91 & 5.07 & 4.72 & -- & -- \\
58519.7 & 5.87 & 5.05 & 4.71 & 4.38 & -- \\
58534.6 & 5.93 & 5.11 & 4.76 & -- & -- \\
58617.4 & 5.81 & 5.01 & 4.67 & 4.32 & -- \\
58630.4 & 5.85 & 5.03 & 4.69 & 4.35 & 4.67 \\
58632.4 & 5.84 & 5.06 & 4.70 & 4.36 & 4.66 \\
58636.4 & 5.90 & 5.04 & 4.70 & 4.38 & -- \\
58653.3 & 6.00 & 5.12 & 4.79 & 4.42 & -- \\
58686.3 & 5.96 & 5.09 & 4.76 & 4.42 & -- \\
58706.3 & 5.86 & 5.04 & 4.71 & 4.37 & -- \\
58721.2 & 5.83 & 5.02 & 4.7 & 4.34 & -- \\
59036.4 & 5.91 & 5.07 & 4.74 & 4.40 & -- \\
59376.4 & 5.93 & 5.12 & 4.77 & 4.43 & 4.72 \\
59451.3 & 5.94 & 5.13 & 4.76 & 4.42 & -- \\
59776.4 & 5.87 & 5.07 & 4.73 & 4.38 & 4.65 \\
\hline
\end{tabular}\\
\end{table*}

In this work we make use of the ultraviolet (UV) spectroscopy obtained by the IUE satellite in 1978--1990 and by the Swift/UVOT in 2015--2023 (Roming et al., 2005). The latter was reduced using the \textsc{heasoft} package (v6.31.1, NASA High Energy Astrophysics Science Archive Research Center (Heasarc), 2014).

\section{PHOTOMETRY ANALYSIS}

Tables~\ref{tab:IR_observation} and \ref{tab:JHKLM_observations} list the IR photometry for T~CrB in the quiet and active state. A large amount of $J$ and $K$ measurements allows us to state that the mean brightness in these bands did not depend on the activity state of the hot component. Then, we can assume that the mean brightness in other bands was constant, too: $\overline{J}=5.91\pm.06$, $\overline{H}=5.07\pm0.05$, $\overline{K}=4.74\pm0.05$, $\overline{L}=4.38\pm0.04$, $\overline{M}=4.68\pm0.05$.

We performed a frequency analysis of the IR measurements presented in Table~\ref{tab:IR_observation} using an upgraded version of the \textsc{l2} program created by Yu.~Kolpakov. The code implements the fitting of a time series by a third-order polynomial, which reconstructs the long-term trend, and then the Fourier analysis up to the third harmonic component of the residuals between the observational data and the polynomial. There are two prominent peaks in the resulting power spectrum of the $J$-band light curve corresponding to the periods $P=227.55\pm0.1$\,d and $P'\approx 0.5 P$. The first one coincides with the orbital period found by Fekel et al. (2000) from the radial velocity curves for the cool component, and with the period found by Tatarnikova et al. (2013) based on the IR-photometry obtained in 1987--2003. We determined a new ephemeris for the times of minimum brightness when the red giant is located between the hot component and the observer ($\varphi=0$): $JD_{min} = 2455828.9 + 227.55 \times E$. 

\begin{figure}
\includegraphics[scale=0.85]{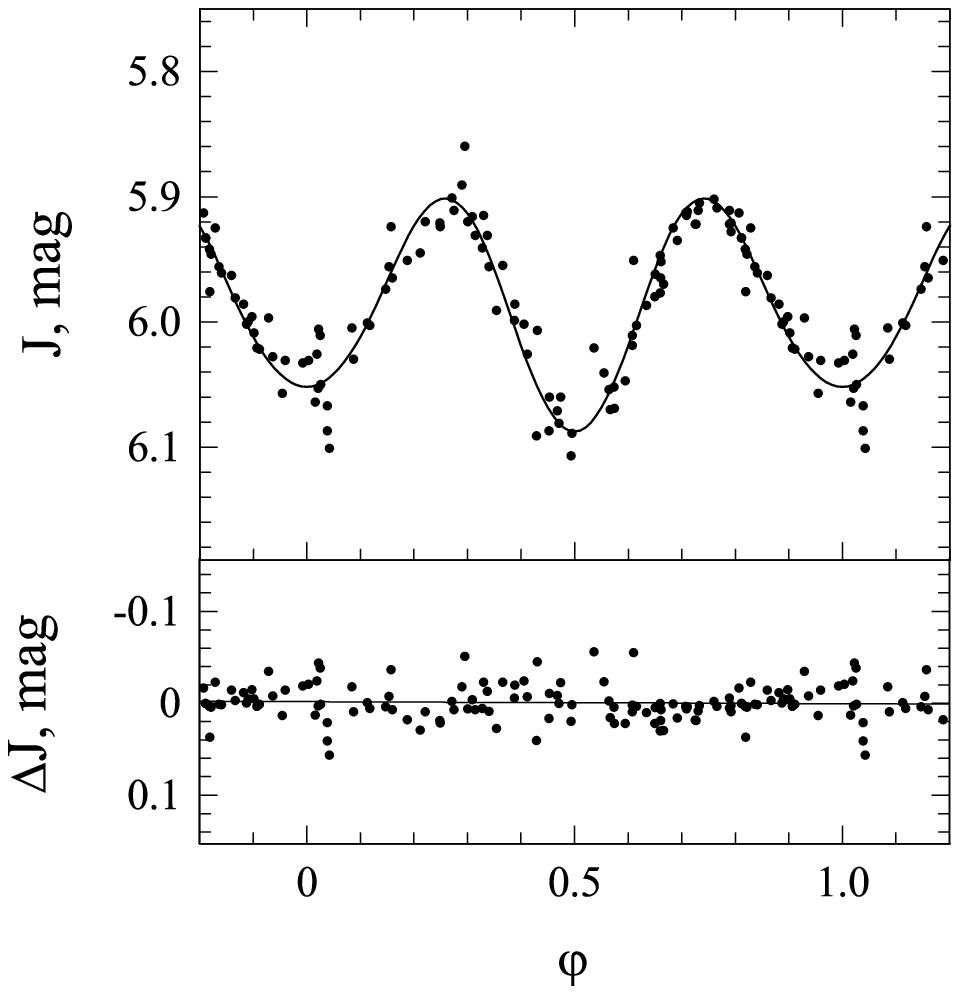}
\caption{The observed $J$ light curve (dots) of T~CrB folded on the period $P=227.55$\,d and model light curve (solid line) of a tidally distorted red giant (the model parameters are $\mu = 1$, $q=0.57$, $i=58^\circ$, $T_{eff}=3500$\,K, see text for details). The bottom panel shows the deviations of the observed magnitudes from the model curve.}
\label{fig:J_model}
\end{figure}

The $J$ light curve folded on the period $P$ is shown in Fig.~\ref{fig:J_model}. The period found from the $K$ light curve is equal to $P$ within error. Table~\ref{tab:IR_observation} demonstrates that a sharp maximum which happened during the active state and fell on April 2016 as was reported by Munari (2023), occurred at phase $\sim0.3$. It is seen in Fig.~\ref{fig:J_model} that the $J$ brightness measured in April 2016 is higher by just $\sim 5$\% than the mean brightness at this phase. So, we can state that the transition of the hot component to the active state in 2015 barely affected the IR brightness of the system (this is also supported by the mean brightness constancy).

The presence of two periods, one of which is equal to the orbital one and the other is $0.5P_{orb}$, points out the ellipsoidal effect when the total brightness varies due to the changing aspect of the tidally distorted star with respect to us, and to the variations of temperature on the visible stellar surface. Shahbaz et al. (1997) investigated this effect for the T~CrB system.

When analysing the IR SED, we can neglect the input from the accretion disc and nebula compared to that from the red giant (Maslennikova et al., 2023). So, we have modelled the $J$ and $K$ light curves of T~CrB considering the orbital motion of the tidally distorted cool component (Tjemkes et al., 1986). The near-IR red giant's SED is well fitted by a black body (Pickles, 1998). According to the data in Table~\ref{tab:IR_observation}, we have $J-K=1.17$ for the mean colour and, after a small correction for interstellar reddening applied, this corresponds to the M3\,III-M4\,III spectral type and the effective temperature $T_{eff}\approx 3500$\,K.

Hachisu and Kato (1999) modelled the optical light curve of T~CrB during the first 300~days of the 1946 outburst. They explained the noticeable secondary light maximum by the reflection of the X-ray radiation from the hot component by the red giant and accretion disc. The modelling took into account the X-ray luminosity (30--100\,\AA) simulated for the best-fitting model of T~CrB ($M_{hot}=1.377\,\textrm{M}_\odot$, X\,=\,0.7, Z\,=\,0.02). With a distance of 1~kpc adopted, the simulated X-ray flux was $\lg\,F_{SSX}=-6.5$ at the moment of X-ray light maximum, and it decreased by not more than two orders of magnitude by the middle of the secondary maximum (if the X-ray light curve is linearly extrapolated to the moment of secondary maximum).

The X-ray fluxes related to the IR light curve simulated here are much smaller. According to Kennea et al. (2009), the dereddened {\it SWIFT/XRT} flux in a wider range of 0.3--10~keV was just $3.8\times10^{-11}$~erg/cm$^{2}$s that is smaller by four orders of magnitude than the maximal theoretical flux in a much narrower range. Later, during the super-active stage, the X-ray flux from T~CrB dropped by another order of magnitude (see Kuin et al. (2023) and references therein) and returned to previous values in 2023. Following the arguments given above, we neglect the reflection effect.  

\begin{figure*}
\includegraphics{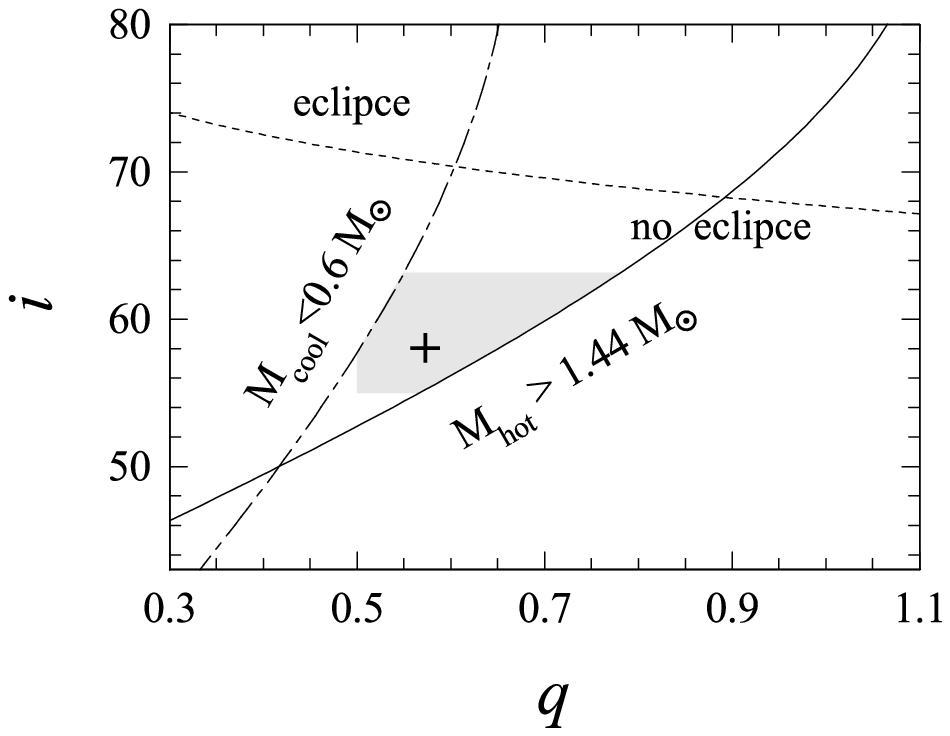} 
\caption{The $(q, i)$ diagram for a fixed value of mass function $f(m)=0.3224$ (Fekel et al.(2000)). The curves demonstrate the constraints corresponding to $M_{cool} > 0.6 M_{\odot}$, $M_{hot}<1.44 \textrm{M}{\odot}$ and the absence of eclipses. Grey area represents the space of possible values where the models that fit the observations fall with a 90 per cent probability (according to the Fisher criterion). The cross indicates the model with the smallest deviation from the observations $q=0.57$, $i=58^\circ$.} \label{fig:limits}
\end{figure*}

When modelling the SED of the cool component, we assumed the surface averaged temperature to be $T_{eff}$, we took into account the limb darkening from Claret (2000) and the gravitational darkening for stars with convective envelopes according to Lucy (1967) assuming the exponent $\beta=0.08$. When modelling the IR light curves, the orbital inclination $i$ is determined quite accurately, whereas the mass ratio $q$ may vary in a large enough interval. To additionally constrain $q$ and $i$ we assumed that the red giant's mass is bigger than 0.6\,$\textrm{M}_\odot$ and the hot component's mass must not exceed the Chandrasekar limit. Besides, as follows from UV observations T~CrB is not an eclipsing binary (Selvelli et al. (1992)). Taking into account the known mass function (Fekel et al. (2000)) and assuming that the cool component is filling its Roche lobe we can significantly constrain the values of $q$ and $i$. This is illustrated by Fig.~\ref{fig:limits}. Based on these considerations and applying the Fisher criterion we found the system parameters with a 90 per cent probability to lie in the intervals $q \in [0.5, 0.77]$, $i \in [55^{\circ}, 63^{\circ}]$. The model that best fits the observations has the following parameters: the Roche-lobe filling factor $\mu=1.0$, the mass ratio $q=M_{cool}/M_{hot}=0.57$, the orbital inclination $i=58^{\circ}$ (see Fig.~\ref{fig:J_model}). Using the mass function $f(m)=0.3224$ (Fekel et al. (2000)) we derived 1.30\,$\textrm{M}_\odot$ for the hot component and 0.74\,$\textrm{M}_\odot$ for the cool companion.     

\begin{figure*}
\includegraphics[scale=0.9]{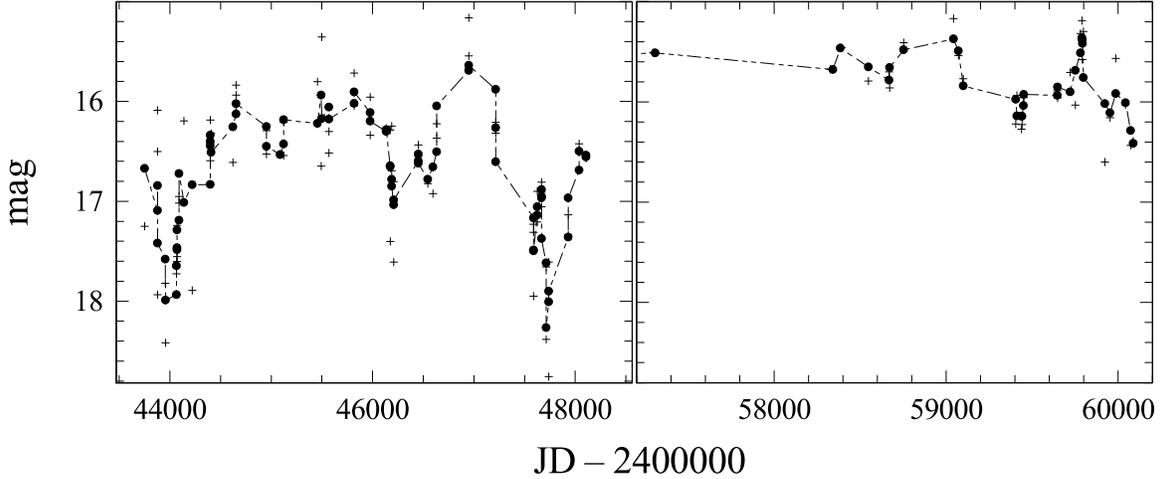}
\caption{UV light curves for T~CrB (in arbitrary magnitudes) reconstructed from the UV IUE (left panel)
and UVOT (right panel) spectra. The crosses indicate individual estimates, the dots correspond to the 3-point running average. See text for details.}
\label{fig:IUE_UVOT}
\end{figure*}

Contrary to the IR spectral range where the radiation from the cool component dominates, a significant brightness variation is observed in the UV domain. T~CrB was observed many times in 1978~--- 1990 with UV spectrographs of the IUE space observatory. The obtained data were barely studied. After the star entered the super-active state in 2015, the UV spectra began to be obtained on the UVOT telescope of the Swift space observatory. Fig.~\ref{fig:IUE_UVOT} shows the UV light curve for T~CrB reconstructed from these spectra. For this purpose, we measured the mean flux in sectors free of emission lines. We used the region centered at $\lambda 1850$\AA{} for the short-wave range of IUE (SWP spectra) and the region centered at $\lambda 2050$\AA{} for the long-wave ranges of IUE (LWR and LWP spectra) and for the UVOT spectra. Analysing the IUE spectra obtained close in time we found that the magnitudes calculated via $mag = -2.5 \lg ( flux)$ are almost equal for the short and long-wave regions. Therefore, we combined all the data to create one light curve.

The combined light curve demonstrates several UV brightenings with the most prominent one occurring during the super-active stage of 2015~--- 2023. A smaller one was observed in 1987 but it did not show up in the visible range according to the AAVSO data. Deep narrow light minima were observed in 1979 and 1989. These seem not to correlate with the relative orbital location of binary components since they occurred at different phases (about 0.8 and 0.5, respectively). The frequency analysis of the UV brightness variations shows that there is no peak at orbital frequency in the power spectrum. A period of $\sim3.3$~yr can be distinguished. Earlier a close period was identified by Ilkiewicz et al. (2016) based on optical data analysis.

T~CrB is one of the first symbiotic stars with detected flickering (Walker, 1957). In order to study flickering, T~CrB is usually observed in the $U$ and $B$ bands where the cool component's input is smaller than in redder wavelength regions. Before the system entered the super-active state in 2015, the flickering amplitude had been $0\,.\!\!^{\rm m}08-0\,.\!\!^{\rm m}16$ in $B$ (Sokoloski et al. (2001), Zamanov et al. (2010)). In the super-active state the amplitude almost did not change and was $0\,.\!\!^{\rm m}08-0\,.\!\!^{\rm m}13$ (Zamanov et al. (2016), Maslennikova et al. (2023), Shore et al. (2023)). When the $B$ brightness started to decrease at the end of April 2023, the flickering amplitude rose up to $0\,.\!\!^{\rm m}26$ Minev et al. (2023). Our $B$-band follow-up photometry carried out on June 8, 2023 simultaneously with the follow-up spectroscopy showed that the flickering amplitude had decreased significantly to a value of $0\,.\!\!^{\rm m}07$.

\section{SPECTROSCOPY ANALYSIS}

\begin{figure*}
\includegraphics[scale=0.8]{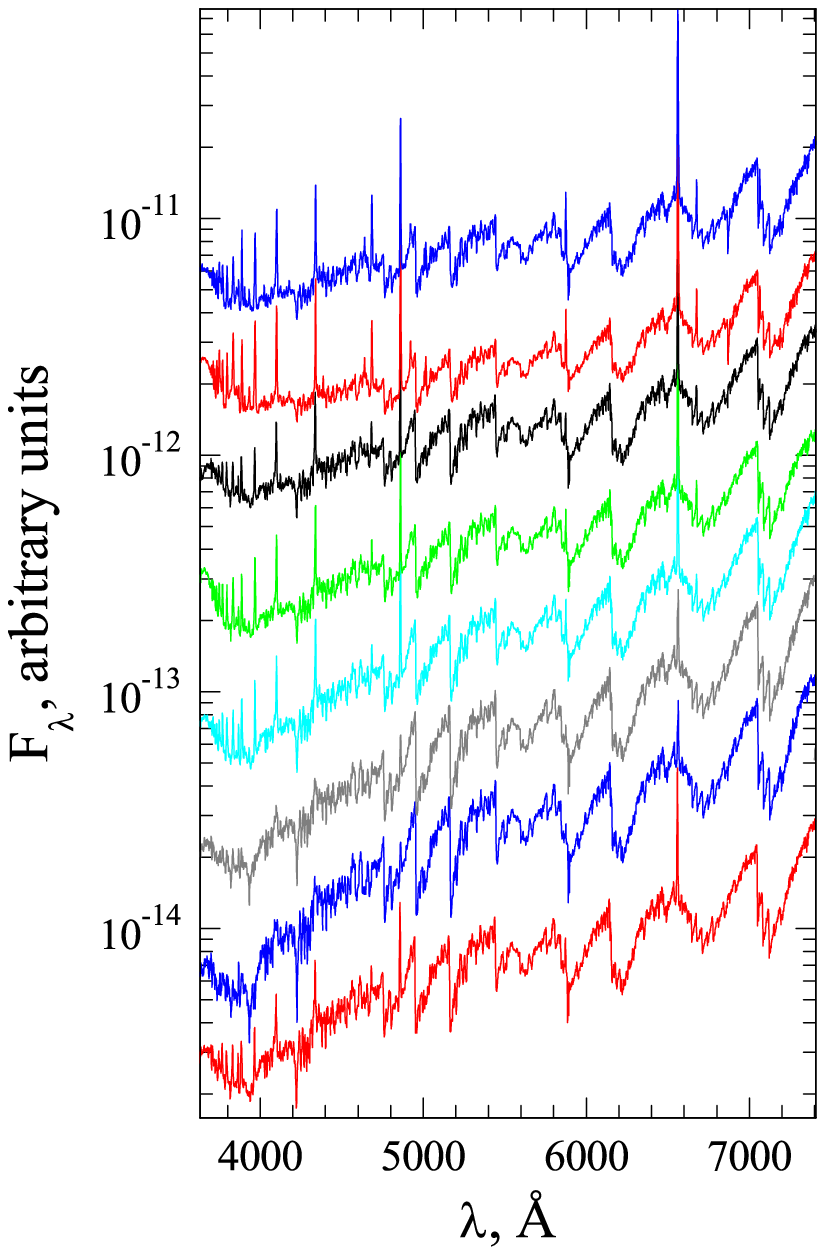}
\includegraphics[scale=0.5]{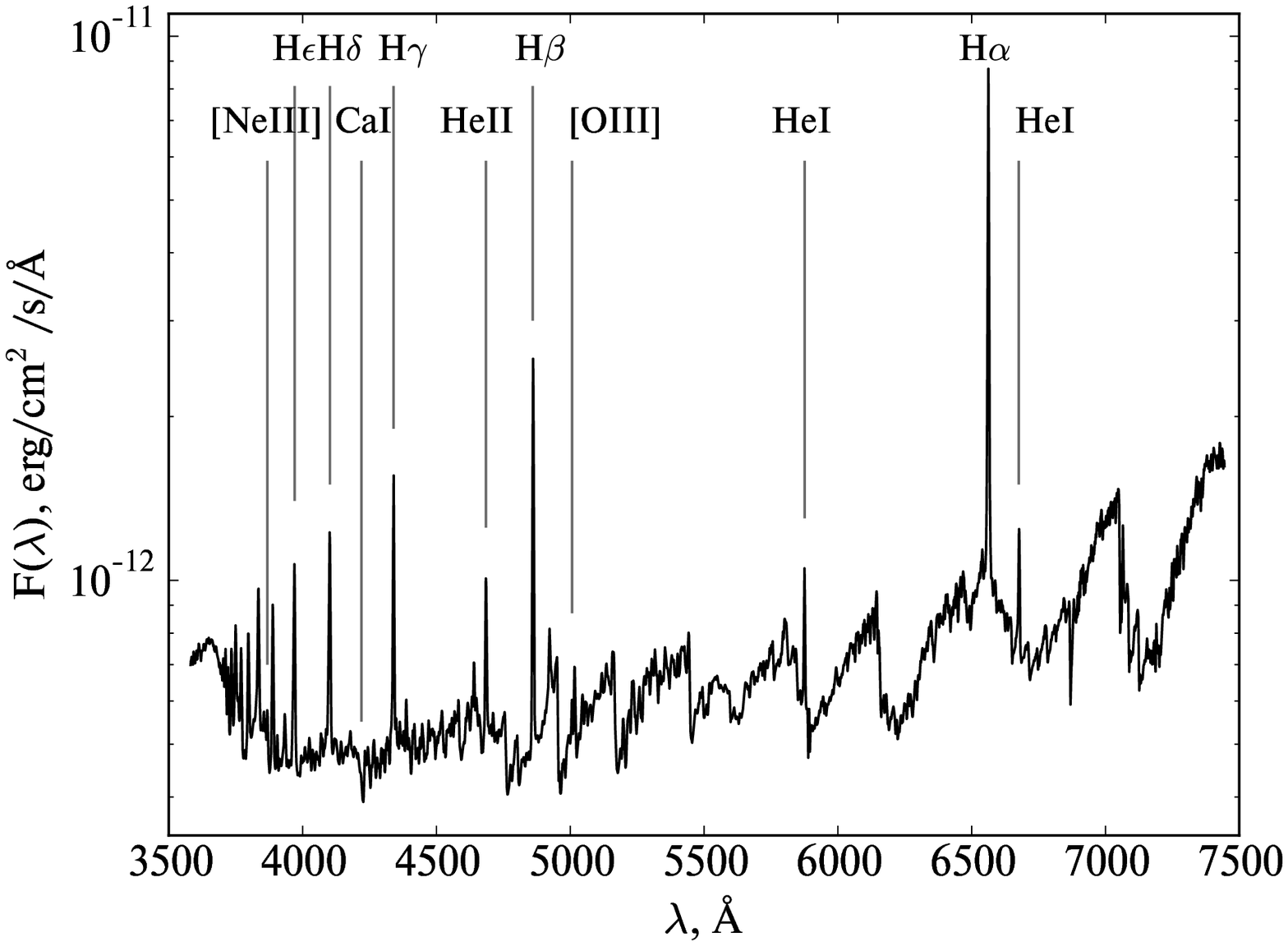}
\caption{The observed spectra for T~CrB (left panel) obtained on (from top to bottom) 22.03.2020, 05.01.2022, 27.01.2023, 26.02.2023, 09.03.2023, 24.06.2023, 26.05.2023, 14.07.2023. The fluxes are multiplied by an arbitrary constant for clarity. The right panel shows the spectrum of T~CrB obtained on 05.01.2022 with the most prominent lines identified.}
\label{fig:All_sp_T_CrB}
\end{figure*}

Fig.~\ref{fig:All_sp_T_CrB} demonstrates the sequence of spectra for T~CrB obtained in 2020~--- 2023 during the current active state decline. Those of the spectra listed in Table~\ref{tab:observation} that are almost identical to the presented ones are not shown in the figure. A significant fading of emission lines and a variation in the slope of the continuum are evident. The emission lines except H$\alpha$ are barely seen in the last two spectra of 2023 whereas the absorption bands and lines (Ca\,I $\lambda$ 4227 and those of the "blue"{} TiO band) have become more prominent. Thus, for example, a Ca~II H absorption is well seen while it was filled by the H$\epsilon$ emission earlier. The super-active stage of T~CrB may be considered to be over in April 2023.

In Fig.~\ref{fig:All_sp_T_CrB} the spectra are vertically arbitrarily shifted for clarity but the flux in the red part of the spectrum where the cool component dominates may be considered nearly constant, varying only slightly due to the ellipsoidal effect. Then we can assume that in May~--- June 2023 the short-wave flux decreased by a factor of 2 compared to 2022~--- early 2023. The radiation input from the nebula and accretion disc significantly dropped. Whereas the [Ne\,III] $\lambda 3869$ line did not change noticeably. It should be noted that the spectra are not arranged chronologically, and, in fact, the periods of emission lines weakening alternate with those of strengthening (e.g., the line fluxes increased again in July 2023 after a period of fading in May~--- June 2023).

\begin{figure}
\includegraphics[scale=1]{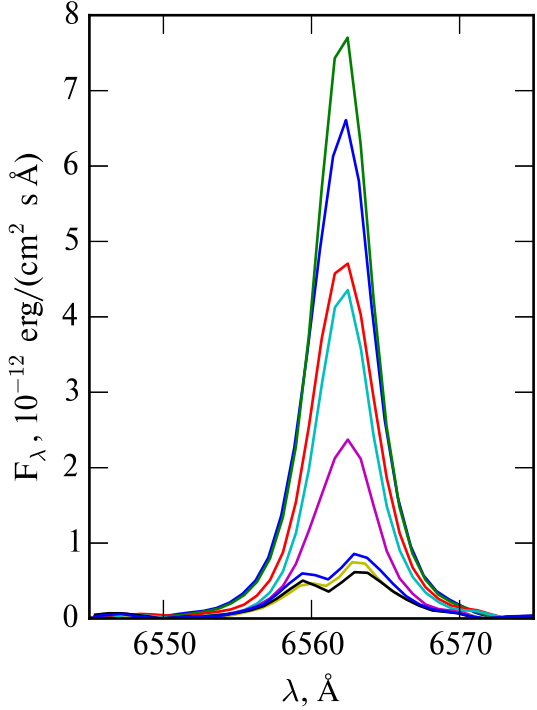} 
\caption{The H$\alpha$ profile in the spectra obtained on (from top to bottom) 05.01.2022, 22.03.2020, 26.02.2023, 09.03.2023, 26.05.2023, 14.07.2023, 24.06.2023 and 04.07.2023.}
\label{fig:TCrB_Halpha}
\end{figure}

Fig.~\ref{fig:TCrB_Halpha} shows the variation of the H$\alpha$ profile observed from early 2020 till July 2023. The maximum line flux was $\sim\,3\times10^{-11}$~erg/cm$^2$/s, the minimum one~--- $3.3\times10^{-12}$~erg/cm$^2$/s. Between 2023 June 8 and 21 (i.e. at phases from 0.84 to, at least, 0.94) the lines began to show a double-component profile. It is worth mentioning that no double-component line structures were observed on 09.12.2022 at $\varphi=0.99$ during the previous cycle. A double-component H$\alpha$ profile can be fitted by adding an absorption with a radial velocity of $-72\pm3$~km/s$^{-1}$. At the same time, the centre of emission corresponds to a radial velocity of $-32\pm5$~km/s$^{-1}$ that is close to the $\gamma$-velocity of the system derived by Fekel et al. (2000) based on absorption spectrum of the cool component. The radial velocity of the additional absorption feature is close to that of the central one observed by Stanishev et al. (2004) at similar phases (note that in Stanishev et al. (2004) not a photometric but spectroscopic ephemeris is used and it is shifted by a quarter of period).    

\begin{figure} 
\includegraphics[width=1\linewidth]{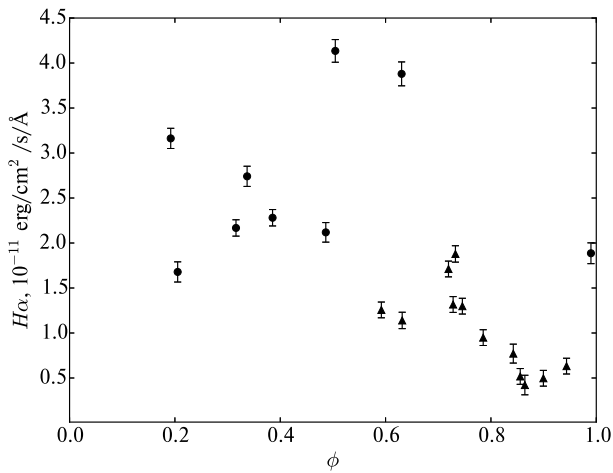} 
\caption{The variation of H$\alpha$ flux as a function of orbital phase based on spectra obtained before the middle of April 2023 (circles) and later (triangles).}
\label{fig:Halpha_phi}
\end{figure}

The Balmer lines may originate in different regions of the symbiotic system (accretion disc, nebula, hot spot). Fig.~\ref{fig:Halpha_phi} shows the variation of H$\alpha$ flux with phase. In the super-active state the H$\alpha$ flux was observed to somewhat decrease near the phase $\varphi=1$. But we should mention that the H$\alpha$ fluxes measured on 24.01.2023 and 27.01.2023 (at phase $\varphi\sim0.2$) differ by a factor of about 2. No significant dependence on phase is seen in Fig.~\ref{fig:Halpha_phi}, and this gives evidence that the H$\alpha$ flux is more sensitive to the activity state of the system than to the orbital motion.

Usually the SED of symbiotic stars is well fitted by a three-component model consisting of a red giant, a hot component and a nebula. By comparing the red-wavelength spectra of T~CrB with averaged spectra of red giant standards taken from Pickles (1998) we classify the spectral type of the cool component as M4\,III and it barely changed during our observations. The temperature of the hot component is $10^5$~K according to Selvelli et al. (1992).

\begin{figure*}
\includegraphics[scale=0.9]{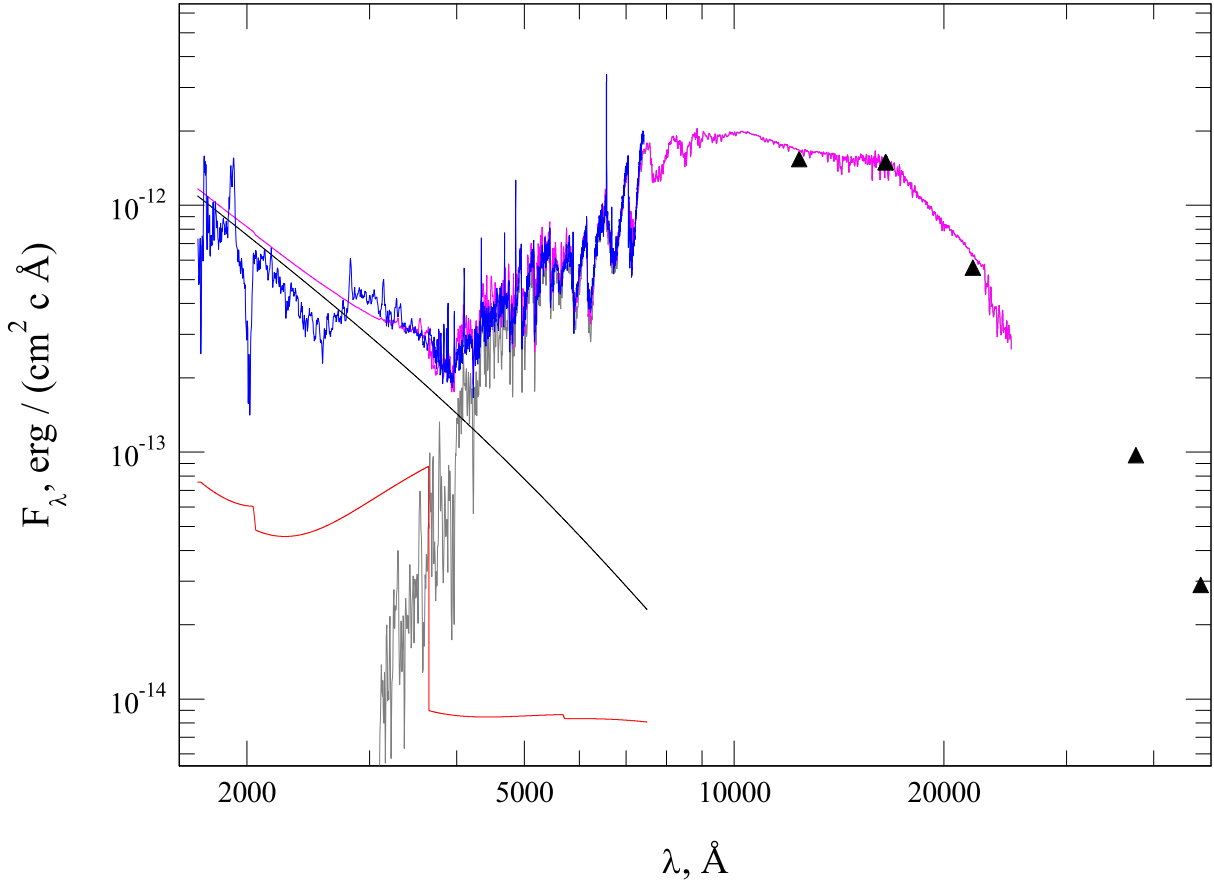}
\caption{The dereddened spectrum of T~CrB (blue line) and the modelled SED of the system (magenta line). The optical spectrum was obtained on 09.12.2022, the UV spectrum (Swift)~--- on 10.12.2022. The triangles show the $JHKLM$ magnitudes for T~CrB obtained on 26.05.2019. The grey line denotes a M4\,III red giant, the red line stands for the combination of a hot component with $T_{eff}=10^5$~K and a nebula with $T_e=10^4$~K, the black line represents an accretion disc when $i=56^\circ$, $\dot M=4 \times 10^{-8} \textrm{M}_\odot/$yr.}
\label{fig:IR_UV}
\end{figure*}

Fig.~\ref{fig:IR_UV} shows a dereddened spectrum of T~CrB obtained during the super-active stage. In the spectrum there are the bands of TiO and the Ca~I $\lambda$ 4227 line in absorption, the following features are well distinguished: the Balmer jump, the emission lines of the Balmer series, He~I ($\lambda$ 5875.6, 6678.1), He~II ($\lambda$ 4685.7), [Ne~III] ($\lambda$ 3869), [O~III] ($\lambda$ 4363, 4959, 5007), Mg~II  ($\lambda$ 2796.3,  2803.5). One can see that the radiation from a so-called "warm"{} component is needed to fit the SED in the blue and UV, and this component is an accretion disc.

We used the equations from Tylenda (1977) to model the SED of the accretion disc:

\begin{displaymath}
F_{disk}(\lambda)=\frac{2hc^2}{\lambda^5 d^2} \sin{i} \int\limits_{R_1}^{R_{out}} \frac{2\pi R}{exp\left ({\frac{hc}{\lambda k T(R)}}\right )-1}\, \mathrm{d}R,
\end{displaymath}
\begin{displaymath}
T(R)= \left[\frac{3GM_1 \dot M}{8\pi\sigma R^3}(1-(R_1/R)^{0.5})\right]^{0.25},
\end{displaymath}
where $R_1$ is the inner disc radius, $R_{out}$~--- the outer disc radius, $d$~--- the distance to the system, $i$~--- the angle between the normal to the disc surface and the line of sight, $M_1$~--- the mass of the hot component, $\dot M$ denotes the rate of accretion onto the hot component. We adopt the inner disc radius $R_1=0.004\,\textrm{R}_\odot$ (Pshirkov et al. (2020)) that is equal to the radius of a $1.3\textrm{M}\odot$ white dwarf (see above). This is the smallest possible value of $R_1$ and it will grow with increase of the magnetic field strength. The orbital inclination for T~CrB is set equal to $i = 58^\circ$ which we have derived through modelling the ellipsoidal effect. We have considered an outer disc radius of $1\,\textrm{R}_\odot$ (Selvelli et al. (1992), Maslennikova et al. (2023)).

\begin{figure*}
\includegraphics[scale=1]{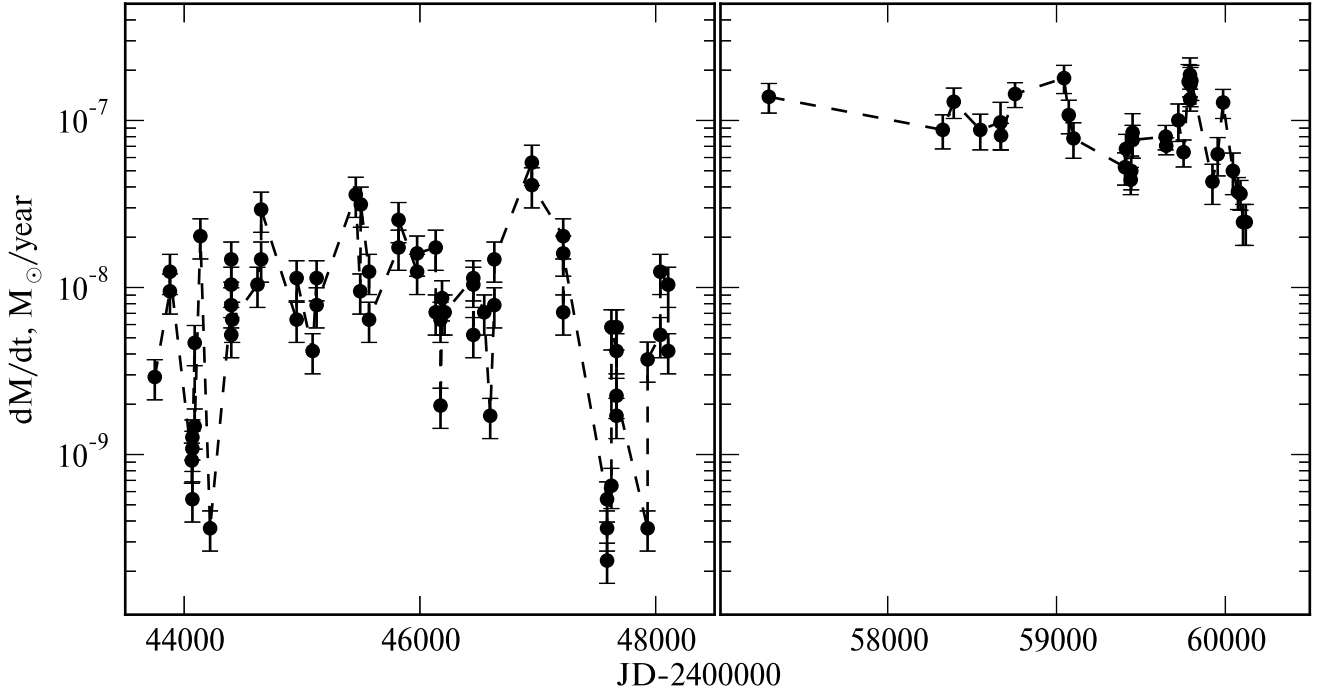} 
\caption{Variation of accretion rate estimated from UV spectra.}
\label{fig:dM}
\end{figure*}

By fitting UV spectra obtained by IUE and UVOT/Swift and optical spectra with the given model we found the accretion rate (Fig.~\ref{fig:dM}) which affects the temperature of the accretion disc. It had been smaller than $6\times10^{-8}\,\textrm{M}_\odot$/yr before the system transitioned to the super-active state. The accretion rate was $4\times10^{-8}$~--- $2\times 10^{-7}\,\textrm{M}_\odot$/yr from 2015 till the middle of April 2023. When the super-active state was over, the accretion rate dropped to $2.5\times10^{-8}\,\textrm{M}_\odot$/yr that approximately coincides with the values derived from IUE spectra. We could vary outer disc radius $R_{out}$ instead of $\dot M$ to explain the UV spectral changes. But to do so we would have to decrease and then increase again the size of the disc by a factor of 6-7 on a time-scale of several tens of days. Note that the inner disc radius $R_1$ barely affects the SED in the UV region.  

The volume emission measure for the nebula in Fig.~\ref{fig:IR_UV} is $1\times10^{58}$~cm$^{-3}$ which is a little smaller than $EM=4\times10^{58}$~cm$^{-3}$ estimated on 25.08.2020 (Maslennikova et al., 2023). The change in Balmer jump demonstrated in Fig.~\ref{fig:All_sp_T_CrB} can be well explained by the decreasing of emission measure since the end of April 2023. And indeed it was $3\times10^{57}$~cm$^{-3}$ on 04.07.2023.

We carried out simultaneous follow-up observations of T~CrB with the TDS spectrograph on the 2.5-m telescope and the CCD photometer on RC600 at the CMO SAI MSU on 08.06.2023.

\begin{figure}
\includegraphics[scale=0.75]{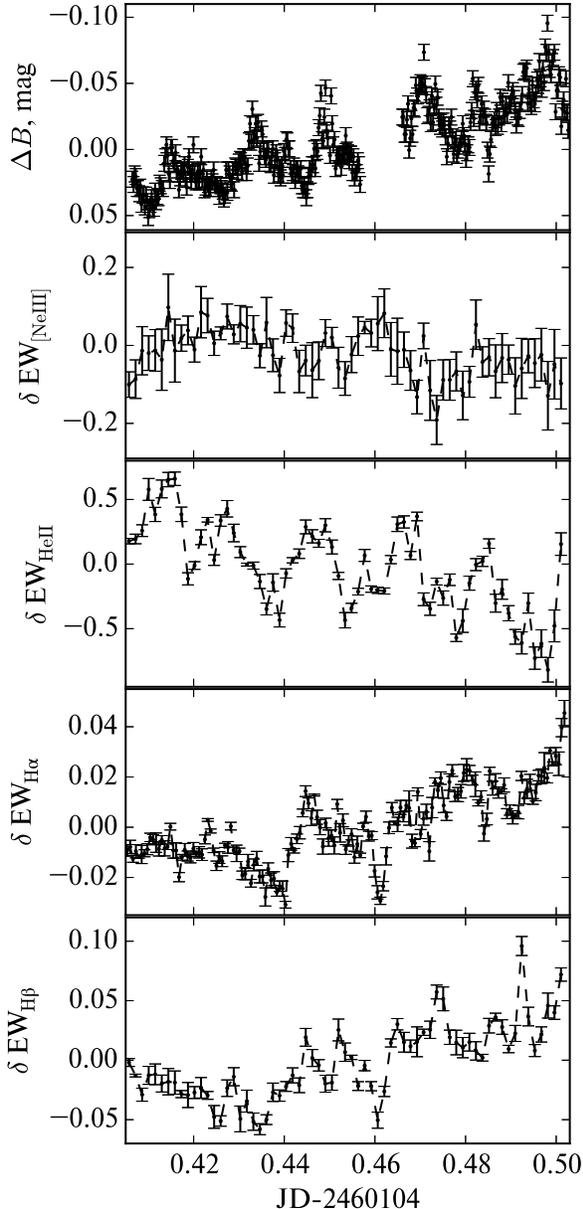}
\caption{The $B$-band light curve and the evolution of equivalent widths of the H$\alpha$, H$\beta$, He\,II $\lambda 4686$, [Ne\,III] $\lambda 3869$ lines derived from the follow-up observations carried out on 08.06.2023.}
\label{fig:Spectral_monitoring}
\end{figure}

The instability of seeing and telescope driving led to the fact that the registered flux experienced changes up to a factor of 2 due to the varying light loss at the slit. As the absolute flux measurements are impossible under these circumstances we searched for flickering in terms of equivalent widths (EW) of emission lines. In order to derive EW of a line, it is necessary to estimate the continuum level near the line but this task is difficult enough because of numerous absorption lines. As we were interested to detect only the variation in EW we did the following. First, to remove the impact from light loss at the slit we normalized the spectra to the integral flux determined as $f_{\lambda}=F_{\lambda}/\int\limits_{\lambda_1}^{\lambda_2}F_{\lambda}{\rm d}\lambda$ for a large enough integration range $\lambda_1-\lambda_2$. The exact size and limits of the range are not important but it must locate near the given line as the light loss may slightly depend on wavelength. Second, to study the variation of spectral lines we considered the difference between an individual spectrum $f_{\lambda}$ and a median one $\overline{f}_{\lambda}$: $D_{\lambda} = f_{\lambda} - \overline{f}_{\lambda}$. This difference appears as noise for invariable spectral features. Whereas the variable lines stand out as meaningful positive or negative residuals of the difference. To obtain a quantitative measure of these residuals we calculated a normalized integral over the line profile $\int D_{\lambda}{\rm d}\lambda/\int (\overline{f}_{\lambda}-f_c){\rm d}\lambda$ which represents the relative variation in the EW of the line $\delta{\rm EW}=\Delta {\rm EW}/{\overline{\rm EW}}$. $f_c$ is the continuum level which needs to be determined only once and its uncertainty enters only $\overline{\rm EW}$ but not the variable component which we aim to study. The error of $\delta{\rm EW}$ was estimated as a standard deviation from zero of six similar integrals with the same size of integration range taken on the left and on the right from the line. In Fig.~\ref{fig:Spectral_monitoring} we show the values of $\delta{\rm EW}$ and their errors for the H$\alpha$, H$\beta$, He\,II $\lambda 4686$, and [Ne\,III] $\lambda 3869$ lines together with the $B$ light curve.

One of our aims of follow-up observations was a search for variations in radial velocity (and therefore we used a narrow slit) but we did not detect systematic variations in the characteristics of line profiles based on differential spectra $D_{\lambda}$.

It is seen from Fig.~\ref{fig:Spectral_monitoring} that the equivalent widths of the Balmer lines demonstrate similar variability which differs from that in other panels. The correlation coefficient between H$\alpha$ and H$\beta$ is larger than 0.6 with no time lag between then. This is supported by the presence of a common feature in the curves observed near the moment 0.46.

The [Ne\,III] $\lambda 3869$ line flux did not show significant variations during our follow-up observations. This fact is in agreement with the result which we found earlier (see Maslennikova et al., 2023) and confirms the conclusion that the line originates in a much more extended region and due to its large size the impact of fast variability of the hot component (and/or accretion disc) is smoothed.

The relative EW of the He\,II $\lambda 4686$ line behaves differently~--- it displays significant variations with an amplitude much larger than observational errors and a characteristic time-scale of $\sim 25$~min. This time coincides with the estimate which we derived earlier (see Maslennikova et al., 2023). The $B$ brightness of the system varies with approximately the same time-scale (see the upper panel in Fig.~\ref{fig:Spectral_monitoring}). But no correlation was found between the $B$ and He\,II $\lambda 4686$ data.

\section{DISCUSSION}

\begin{figure*}
\includegraphics{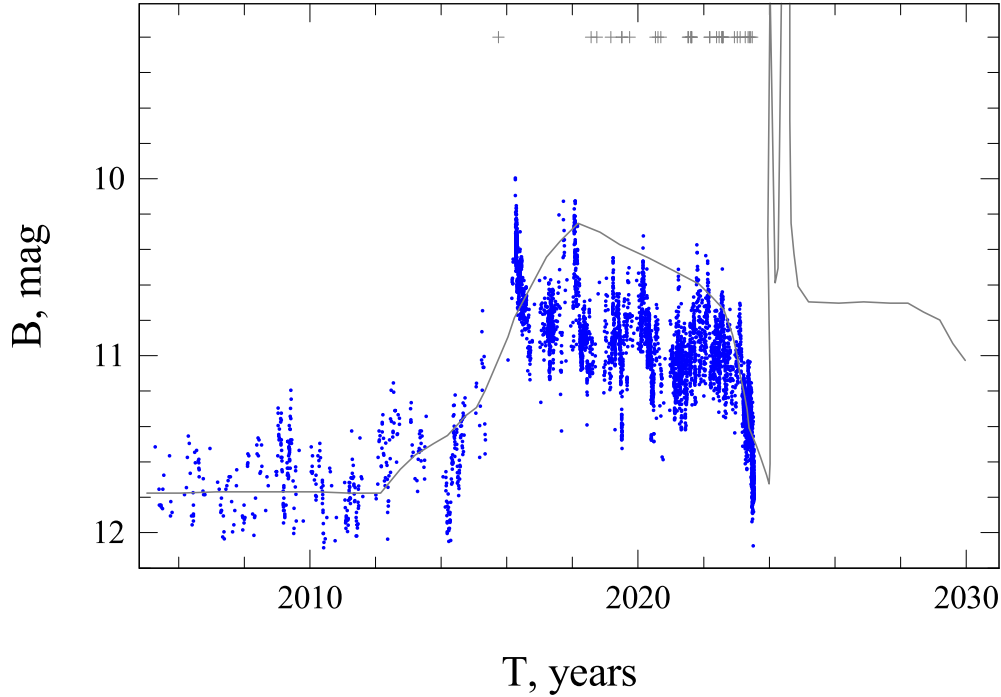}
\caption{The $B$-band light curve of T~CrB composed of AAVSO data (dots). The black line represents the averaged light curve of the 1946 eruption of T~CrB from Schaefer (2023), the crosses indicate the moments of Swift-UVOT observations.}
\label{fig:AAVSO}
\end{figure*}

Fig.~\ref{fig:AAVSO} shows the $B$-band light curve for T~CrB composed of AAVSO data from early 2005 till the middle of July, 2023 (Kloppenborg, 2023). We also plot a template of brightness variation created by Schaefer (2023) based on $B$-band photometry of the 1946 outburst. It is clearly seen that the current "high"{} state develops with a similar characteristic time-scale but smaller amplitude than the super-active state of 1938-1946. As the accretion disc provides the main input to the $B$ brightness of the system we can assume that the pre-outburst accretion rate is now smaller that was previously. Since the middle of March, 2023, T~CrB has been exhibiting a characteristic fading episode which is also present in the outburst template. So, the observed light curve matches the mean light curve of the 1946 eruption very well. If we assume that the upcoming outburst will follow the 1946 scenario, we can expect that the eruption will occur in the beginning of 2024.

The UV data obtained by the Swift satellite also indicate the presence of growing activity (see Fig.~\ref{fig:IUE_UVOT}). The mean UV flux at $\sim 2000$\AA{} measured during the active state is larger by a factor of 2-2.5 than the mean flux measured by the IUE satellite in 1979-1990. If we assume that the change in UV flux is due to the variation of accretion rate, then the latter has to vary from $< 10^{-9}\textrm{M}_\odot$/yr for the minimum observed fluxes to $\sim 10^{-7}\textrm{M}_\odot$/yr for the maximum ones. Another factor that can affect the observed UV flux is the changing matter density in the line of sight which leads to the change in extinction. According to Kuin et al. (2023), the X-ray flux behaved in the opposite way~--- it decreased by a factor of 4 during the transition to the active state in 2015 and then recovered to previous levels in 2023 when T~CrB was completing the super-active stage.

\begin{figure}
\includegraphics[scale=0.45]{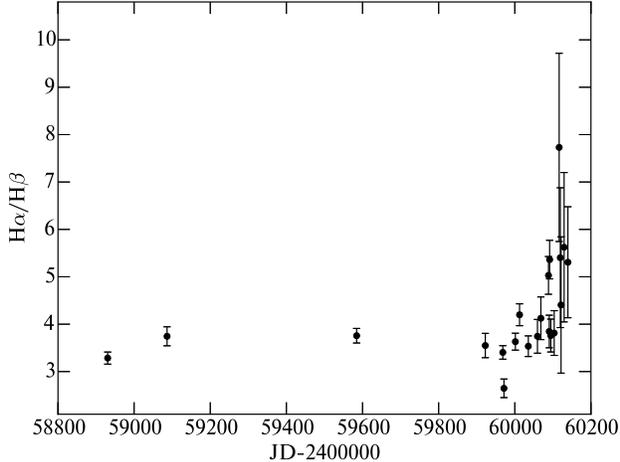} \caption{The variation of the H$\alpha$/H$\beta$ ratio measured from the dereddened optical spectra of T~CrB listed in Table~\ref{tab:observation}}
\label{fig:dm_dt}
\end{figure}

Our spectroscopic observations carried out in 2020-2023 demonstrate that the accretion disc contribution  to the total continuum flux is decreasing. The emission-line spectrum of T~CrB has changed, too: the He\,II $\lambda 4686$ line is barely seen, the Balmer and He\,I lines have weakened significantly, the [Ne\,III] lines are weak but still present in the spectrum. In the case of T~CrB the complex structure of continuum emission does not allow making flux measurements for weak lines. Therefore, the only quantity we managed to measure with high enough accuracy for all nights, is the H$\alpha$/H$\beta$ flux ratio (Fig.~\ref{fig:dm_dt}).

As it follows from Fig.~\ref{fig:dm_dt}, the Balmer decrement (BD) measured in the super-active state (when the accretion rate is high) is consistent with the standard case B recombination values (Osterbrock, 1989). The maximum value of BD $\sim 8$ was measured on 21.06.2023. On the same date a minimum of UV continuum flux was observed. So, we can point out a negative correlation of BD and UV continuum flux which in turn depends on accretion rate (see Fig.~\ref{fig:dM}). The negative BD~-- UV-flux relation was noted for active galactic nuclei and may be due to various reasons (Shapovalova et al. (2019), Wu et al. (2023)). One of the most frequently invoked explanations is the presence of additional extinction near the region where the lines originate. In the case of T~CrB the colour excess needs to be $E(B-V)\sim1$ (assuming a normal extinction law). But we see no evidence for excessive reddening in the SED of the red giant and accretion disc. We suggest that in the case of T~CrB some other mechanism might be responsible for the negative correlation. The study by Gaskell and Ferland (1984) showed that the H$\alpha$/H$\beta$ ratio is highly dependent on the shape of the X-ray to UV continuum. As the relative contribution of the X-ray continuum to photoionization is increased, the amount of free electrons with energies high enough to provide collisional excitation of the third level from the ground state grows. A corresponding increase in X-ray flux accompanied by a decrease in UV and optical flux was also detected for T~CrB.

We carried out 2-h simultaneous photometric and spectroscopic follow-up observations of T~CrB on 08.06.2023 during the decline of the super-active state of 2015-2023. We detected the EW variability of the H$\alpha$ and H$\beta$ lines, and of the He\,II $\lambda 4686$ line, too. In contrast to our similar observations performed on 25.08.2020 and 06.09.2020 during the super-active stage (Maslennikova et al., 2023), we did not detect a time lag between the lines and between the lines and continuum (i.e., $B$-band photometry). Nevertheless, the new time-scales of variations in $B$ and He\,II line are equal to those obtained earlier ($\sim 25$~min). The amplitude of $B$ brightness variations ($\approx 0\,.\!\!^{\rm m}07$) appears similar to that obtained in 2020, too. It contradicts the report by Minev et al. (2023) that the amplitude of flickering increased and recovered to the quiet-state value by May, 2023 when the super-active state of T~CrB was over.

We performed a frequency analysis on new IR photometry obtained in 2011-2023 and refined the ephemeris of T~CrB. The derived period $P=227.55$\,d is orbital and it has not changed since 1958 (Kraft, 1958). A comparison of mean brightness before and during the super-active state shows that different activity states of the hot component barely affect the shape of light curves. This fact allows us to model IR light curves not taking into account the reflection effect which has to be small also because of small X-ray fluxes observed (see the Section 'Photometry analysis'). Fig.~\ref{fig:J_model} demonstrates that the observed light curve can be well fitted only with ellipsoidal effect. So, contrary to Shahbaz et al. (1997), we did not need to invoke an additional cool spot on the red giant's surface to explain a deep minimum near phase $\varphi=0.5$. The parameters of the system derived from modelling are in good agreement with the values published previously by Stanishev et al. (2004) and Tatarnikova et al. (2013): the Roche lobe filling factor $\mu=1.0$, the binary inclination $i=58^\circ$, the mass ratio $q=M_{cool}/M_{hot}=0.57$ (for the model with the least sum of squared residuals). It should be noted that the Roche lobe filling factor strongly affects the depth of minima and is determined with high accuracy, so, we can state that the cool component is filling its Roche lobe.

\section{CONCLUSION}

1. We present the results of photometric and spectroscopic observations of T~CrB obtained in a wide wavelength range from 2011 till 2023. The near-IR photometry points out the ellipsoidal effect with  large amplitude $\Delta J=0.17$ (see Fig.~\ref{fig:J_model}). Based on these data, we derived an ephemeris for the times of primary light minima (when the red giant is located between the hot component and the observer) $JD_{min} = 2455828.9 + 227.55 \times E$, which is in agreement with the results obtained by Fekel et al. (2000) from the radial velocity analysis. The modelling of near-IR light curves with the ellipsoidal effect taken into consideration allowed us to derive the parameters of the binary system: the Roche lobe filling factor for the cool component $\mu=1.0$, the mass ratio $q=M_{cool}/M_{hot}\in[0.5, 0.77]$, the binary orbital inclination $i\in[55^\circ,63^\circ]$. The model that best fits the observations provides the mass of the hot component of 1.30~$\textrm{M}_\odot$ and the mass of the cool component of 0.74~$\textrm{M}_\odot$ if we adopt the mass function of $f(m)=0.3224$ (Fekel et al. (2000)) for the cool component.

2. Based on spectroscopic data obtained in 2020-2023, we detected a considerable change in the H$\alpha$/H$\beta$ flux ratio~--- from $\sim 3$ up to $\sim 8$. We associate this variation with an observed change in the X-ray flux (Kuin et al., 2023) and switching between various mechanisms of excitation of hydrogen atoms. We show that the H$\alpha$/H$\beta$ ratio depends on the rate of accretion onto the hot component (see Fig.~\ref{fig:dM} and Fig.~\ref{fig:dm_dt}).

3. Based on follow-up observations of 08.06.2023 we detected fast variations in the $B$-band brightness with an amplitude of $\sim 0\,.\!\!^{\rm m}07$, in the He\,II $\lambda 4686$ line with a time-scale of $\sim 25$~min and in the H$\beta$ and H$\alpha$ lines, too. The relative changes in EWs were up to 40 per cent. We found no time lag between different lines and $B$-band flux which we had observed for T~CrB earlier (see Maslennikova et al., 2023).

4. The spectrum of T~CrB obtained on 09.12.2022 and combined with the Swift spectrum obtained on 10.12.2022 was fitted by a model SED composed of radiation from a standard M4~III red giant, continuum emission from an accretion disc, a nebula with $T_e=10^4$~K and a hot component with $T_{eff}=10^5$~K. This enabled us to estimate the accretion rate: $\dot M = 4 \times 10^{-8} \textrm{M}_\odot/$yr (assuming that the inner radius of the accretion disc is equal to the radius of white dwarf $0.004~\textrm{R}_\odot$ and the outer one is $1~\textrm{R}_\odot$).

5. A comparison of the AAVSO light curve for the 2005-2023 period with the template of the 1946 outburst created by Schaefer (2023) makes it possible to predict the date of the upcoming classical nova type eruption of T~CrB~--- the beginning of 2024.

\section*{ACKNOWLEDGEMENTS}

This study was performed by using the equipment purchased through the funds of the Development Program of the Moscow State University. The work of A.V.~Dodin (initial reduction and calibration of spectra), A.M.~Tatarnikov (reduction and analysis of UV and IR observations) and N.A.~Maslennikova (data reduction and analysis of high-speed photometry, spectral modelling) was supported by Russian Science Foundation (grant 23-12-00092). We acknowledge with thanks the variable star observations from the AAVSO International Database contributed by observers worldwide and used in this research. We thank the INES archive for providing access to the IUE data. This study has made use of the Swift data provided by the Space Science Data Center (ASI). The authors thank the anonymous referees for carefully reading the paper and providing very useful comments that have contributed to improving the quality of the manuscript.

\section*{CONFLICT OF INTEREST}

The authors declare that there is no conflict of interest.

\end{document}